\documentclass[reprint,secnumarabic,amssymb,amsmath,aip,cha,groupedaddress,frontmatterverbose]{revtex4-1}
\usepackage{graphicx}% Include figure files
\usepackage{dcolumn}% Align table columns on decimal point
\usepackage{bm}
\usepackage[latin1]{inputenc}
\usepackage{natbib}
\usepackage{amsmath}
\usepackage{amssymb}
\usepackage{amsfonts}
\usepackage{dsfont}
\usepackage{xcolor}
\begin{document}
%\begin{footnotesize}
%\preprint{}

\title{Photo-induced spin filtering in a double quantum dot}


\author{J. W{\"{a}}tzel}
\author{A. S. Moskalenko}
\email{andrey.moskalenko@physik.uni-halle.de} \altaffiliation[Also
at ]{A.F. Ioffe Physico-Technical Institute, 194021 St.
Petersburg, Russia}
\author{J. Berakdar}
%\thanks{}
\affiliation{Institut f\"ur Physik, Martin-Luther-Universit\"at
 Halle-Wittenberg,  06099 Halle, Germany}
%\affiliation{}


\date{\today}

\begin{abstract}
We investigate the spin-{dependent} electron dynamics in a double
quantum dot  driven by {sub-picosecond} asymmetric electromagnetic
pulses. {We show analytically that applying
the appropriate pulses, specified here,
 allows a spin separation  on a femtosecond time scale in the sense
that states with a desired spin projection are  localized mainly on
one  of the dots. It is shown how to maintain in time this photo-induced spin-dependent filtering.}
\end{abstract}

%\comment{Remove tick labels on the right hand side of Figs.
%4,5,6,7,9}


\begin{flushright} 1 \end{flushright}

%\keywords{}
\maketitle


Studies of the electron dynamics in quantum well structures triggered by
time-dependent electromagnetic fields revealed a wealth of
interesting phenomena. The driving field can be a continuous wave
(cw) laser, which can be employed to demonstrate processes such as
the coherent suppression of
tunneling~\cite{Grossmann1991,Bavli1993} or the low-frequency
generation~\cite{Bavli1993} in  symmetric double quantum well
(DQW) systems. Coherent control schemes can be used for trapping
of electrons in different quantum wells~\cite{Bavli1993}
%, control of
%electron transfer~\cite{Mancal2001} \comment{this reference has no relation to QWs}
or driving transient current bursts due to quantum
interferences \cite{Dupont1995}.

%Henriksen1999, etc

Another possibility to  control  the electron dynamics on the
subpicosecond time scale is to utilize
ultrashort highly asymmetric electromagnetic pulses
\cite{You1993,Jones1993,You1997,Guertler2003}. Such pulses with
the electric field consisting of an ultrashort and strong
oscillation half-cycle followed by a weak and much longer
oscillation tail of the opposite polarity are called half-cycle
pulses (HCPs). If the amplitude of the tail is small enough it
 hardly  influences the dynamics. The HCP can be considered
then as an unipolar electromagnetic pulse with a small duration
$\tau_{\rm d}$ given by the duration of its first ultrashort and
strong half-cycle.

%The action of a cw laser pulse and a HCP on
%physical systems are fundamentally different. While the
%interaction with a cw laser field is characterized by a single
%frequency and a broad time range, there are generally frequencies
%in a broad range which are delivered to the system at a particular
%time moment in the case of a HCP.
%If $\tau_{\rm d}$ is very small in comparison to the characteristic time $t_{\rm c}$ of the driven
%system the interaction between the pulse and a charge carrier can
%be viewed classically as a kick received  by the charge carrier
%with a strength $\Delta p=\int F(t){\rm d}t$, where $F(t)$ is the
%force induced by the corresponding electric field
%\cite{Tiekling1995,Dion2001,Alex_Molecules2003,Alex_QW2004,Alex_Europhysics2005,Alex_PRL2005}.
Quantum mechanically, the interaction between a HCP and a charge
carrier can be approximately described as a transformation of the
carrier wave function $\Psi(p)\mapsto\Psi(p+\Delta p)$ in the
momentum representation, while in the coordinate representation we
have $\Psi(x)\mapsto\Psi(x)e^{-i\Delta p x/\hbar}$. It is possible
to localize a charge carrier on the subpicosecond time scale by
applying HCP sequences~\cite{Alex_QW2004}.
%In previous
%studies we demonstrated how HCPs may affect the spin degrees of
%the freedom \cite{zhu10,zhu102,zhu1021} in semicondutor
%nanostructures and and pseudo spin in graphene \cite{andrey}.
Here
we deal with the question of
 whether it is possible to separate spin states using ultrashort light pulses, an issue  which
is of relevance for applications in ultrafast spintronics and
quantum computing
devices~\cite{Allwood2002,Loss1998,Kane1998,Zutic2004}. The effect
of the intrinsic spin-orbit coupling on the light-driven charge
and spin dynamics in a DQW system has been investigated recently.
It has been shown that the charge distribution dynamics induced by
applied HCPs become spin-dependent~\cite{Sherman}. Below we show
that the spin-dependent localization of the charge carriers can be
{achieved} by using the effect of the field-induced spin-orbit
coupling
\cite{Andrada1994,Andrada1997,Zhang2006,Gvozdic2006,Ekenberg2008,Zhu2010}.
Because of this field-induced coupling an appropriately directed
HCP can generate a spin-dependent coordinate transfer for a free
carrier confined to one dimension so that we have
$\Psi_{\uparrow}(x)\mapsto\Psi_{\uparrow}(x+\Delta x)$ for the
spin-up state $|\uparrow\rangle$ whereas
$\Psi_{\downarrow}(x)\mapsto\Psi_{\downarrow}(x-\Delta x)$ for the
spin-down $|\downarrow\rangle$ state. In the case of an electron
confined to two coupled quantum dots we will show how to apply
appropriate HCPs to control coherently the electron position depending
 on its spin degree of freedom. The required photon
fields are feasible. Nowadays, light pulses with durations in the
range of
femtoseconds~\cite{Zewail1988,Sell2008,Junginger2010,Krauss2010}
and even attoseconds~\cite{Agostini2004,Corkum2007,Krausz2009} can
be generated.

\begin{figure}[t]
\centering
\includegraphics[width=8.4cm]{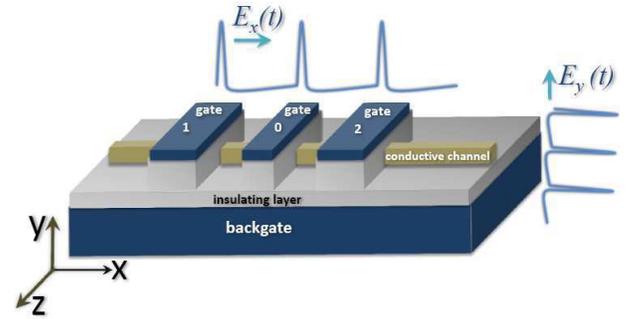}
\caption{(Color online) Schematic representation of the system
considered in this work.   In a quasi-one-dimensional conductive
channel,  two  quantum dots are created and are controllable by
 two local depletion gates: ''gate 1'' and ''gate 2''.  A further ''gate 0'' tunes the tunnel coupling between the dots.
 Shaped, linearly polarized electromagnetic pulses  $E_x$ and $E_y$ are applied. The polarization axis of
 $E_x$ is along the conductive channel, taken as the $x$-axis, whereas the polarization of $E_y$ is along the growth direction
 of the semiconductor heterostructure, chosen as the $y$-axis. }
\label{fig:system1}
\end{figure}

We consider a one-dimensional symmetric DQW, which can be
realized, e.g., on the basis of semiconductor heterostructures and
can be viewed as a double quantum dot (DQD) system. The geometry
of the system and the applied electric field pulses are
illustrated in Fig.~\ref{fig:system1}. The left (right) quantum
dot is located in the region $x<0$ ($x>0$). Due to the quantum
confinement with appropriately high energy barriers in the $y$-
and $z$-directions, the carrier motion is confined only to the
$x$-direction. The control scheme is based on applying two types
of HCPs, which are polarized either along the $x$- or along the
$y$-axis, respectively. Our target is  to  steer spatially the
carrier  depending on its spin state, in particular on its spin
projection to the $z$-axis. We note that both QDs
experience the same pulse field.

The system Hamiltonian is given by
%\begin{equation}\label{for:ham}
$H=H_{0}+H_{\rm K}(t)+H_{\rm SO}(t),$
%\end{equation}
where $H_{0}$ is the Hamiltonian of a free electron in the
DQD. The interaction of the electron with the electric field
$E_x(t)$ that is polarized along the $x$-axis,  is given by
%\begin{equation}\label{kick}
$H_{\rm K}(t)=x~e E_x(t),$
%\end{equation}
where $e$ is the elementary charge. HCPs polarized in this
direction are called here $x$-HCPs. The electric field  $E_y(t)$
of the corresponding $y$-HCPs, polarized along the $y$-axis,
induces the field-dependent Rashba-type spin-orbit
interaction~\cite{Andrada1994,Andrada1997,Zhang2006,Gvozdic2006,Ekenberg2008},
%\begin{equation}
$H_{\rm SO}(t)=\frac{e}{\hbar}\alpha_{_{\rm
SO}}E_y(t)p_{x}\sigma_{z},$
%\label{for:SO}\end{equation}
for the electron in the DQD. Here $\alpha_{_{\rm SO}}$ is the
coupling coefficient, $p_x$ is the component of the momentum
operator along  the $x$-axis and $\sigma_{z}$ is the  Pauli
matrix. We have omitted the term proportional to $p_{z}\sigma_{x}$
in the expression for $H_{\rm SO}(t)$  because the motion in the
$z$-direction is restricted by the carrier confinement. The
electric field pulse shape can be selected, e.g., as
$E(t)=E_0\sin^2(\pi t/\tau_{\rm d})$ for $0<t<\tau_{\rm d}$ (the
field is zero outside of this time range), where $E_0$ is the
electric field amplitude and $\tau_{\rm d}$ is the HCP duration.
The pulse is centered at $t=t_{p}=\tau_{\rm d}/2$ when the
electric field reaches its maximum.

In our treatment we may ignore effects of elastic scattering and
electron-phonon interaction because they take place on much longer
time scales compared with  the times  relevant for this
study \cite{Bastard_book,Cardona_book}. If the duration $\tau_{\rm d}$ of the
 HCP is much shorter than the
characteristic time $t_{\rm c}=2\pi/\omega_{\rm{c}}$
($\omega_{\rm{c}}$ is the frequency corresponding to the energy
difference between the ground and first excited states of the
field-free system) of the undriven system, the impulsive (sudden)
approximation (IA)
\cite{Henriksen1999,Yoshida2000,Daems2004,Matos_Indian_paper} can
be applied when solving the general time-dependent Schr\"{o}dinger
equation. In the framework of the IA the electron interaction with
a $x$-HCP  can be written as
\begin{equation}\label{for:ApprKick}
  H_{\rm K}(t) = x \Delta p \delta(t-\mbox{$t_{p}$})\:,
\end{equation}
where $\delta(x)$ denotes the Delta-function and $\Delta p$
corresponds to the amount of the transferred momentum, which is
given by $\Delta p =e\int\limits_{-\infty}^{\infty} E_{x}(t) {\rm
d}t$. In the case of the sine-square pulse shape field we have
$\Delta p=eE_0\tau_{\rm d}/2$. For the Hamiltonian of the electron
interaction with a $y$-HCP the IA leads to
\begin{equation}\label{TrainApprRashba}
   H_{\rm SO}=p_{x}\sigma_{z}\Delta x\delta(t-t_{p}),
\end{equation}
where $\Delta x =\frac{e}{\hbar} \alpha_{_{\rm SO}}
\int\limits_{-\infty}^{\infty} E_y(t){\rm d}t$ accounts for a
sudden coordinate transfer to the electron. For the sine-square
pulse shape we have $\Delta x=\frac{e}{2\hbar}\alpha_{_{\rm SO}}
E_0\tau_{\rm d}$.
%\footnote{For the joint validity of both
%equations, Eq.~\eqref{for:ApprKick} and
%Eq.~\eqref{TrainApprRashba}, in the IA it is important to assure
%no or small time overlap between the applied $x$- and $y$-HCPs.}

{The DQD parameters are chosen }such that the two lowest
energy levels are well separated{ from the other  states}.
%Hence,
%for a certain range of the pulse parameters the system behaves  basically
% as a two-level system. As a consequence,
{Hence, a two-level
system approximation (TLSA) including the spin can be applied.
 The effective strengths
of the $x$-HCP} and the $y$-HCP can be characterized by the
dimensionless parameters $\beta=x_{12}\Delta p/\hbar$ and
$\alpha=p_{12}\Delta x/\hbar$, respectively. Here
$x_{12}=\langle1|x|2\rangle$ and
$p_{12}=\langle1|p_{x}|2\rangle/i$ are real coordinate and
momentum matrix elements. \footnote{We select the wave functions
to be real.}

\begin{figure}[t!]
\centering
\includegraphics[width=8.2cm]{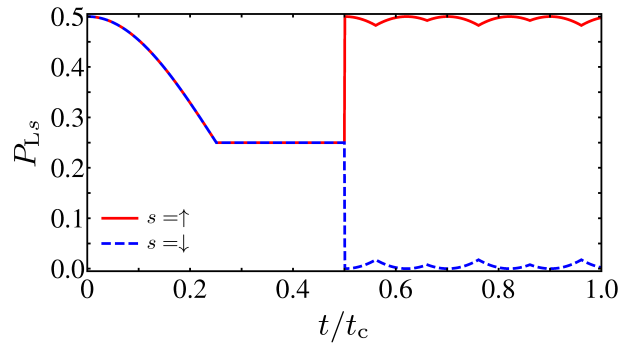}
\caption{(Color online) Time-dependent probabilities to find a
spin-up $P_{L\uparrow}(t)$ (red color) and to find a spin-down
$P_{L\downarrow}(t)$ (blue color) electron in the left well.
Parameters of the excitation by HCPs are
The $x$-HCP is applied at $t_{0}=0.25t_{\rm c}$ and the
$y$-HCP is applied at $t_{0}=0.5t_{\rm c}$ followed by the
periodic $x$-HCP train.
%given in the text.
% For
%the discussed GaAs-based DQD $t_{\rm c}=0.67$~ps.
%At
%$t_{1}=0.25$ the ground state will be finally generated by an
%x-HCP. At $t_{2}=0.50$ the spin separation through an y-HCP takes
%place. It follows the maintenance of that separation for a desired
%period of time, while the system is driven by an appropriate train
%of x-HCPs.
} \label{fig2}
\end{figure}
{The time evolution depends strongly  on the initial
conditions. Two situations are considered:} (a) the
initial condition, which corresponds to an electron being
completely localized in the left well (tunneling initial
condition) and a zero average spin along the $z$-axis, i.e.
$\langle \sigma_z\rangle=0$, and (b) the initially delocalized
state (optical initial condition) with the same spin properties,
which belongs to the ground state of the system. In previous
studies it was shown that a $x$-HCP with appropriate pulse
parameters preceded by a short period of free propagation can be
used to create the optical initial condition from the tunneling
one on an ultrafast timescale \cite{Alex_QW2004}.
%There is, of
%course, no spin-polarization because the excitation is not
%spin-dependent.
A spin polarization can be created if we then
apply a $y$-HCP at the time moment just after the optical initial
condition was created.  {To maintain} the spin polarization an appropriate
periodic train of $x$-HCPs can be used, in analogy to the
maintenance of spin-unpolarized states \cite{Alex_QW2004}.

\begin{figure}[t!]
\centering
\includegraphics[width=8.4cm]{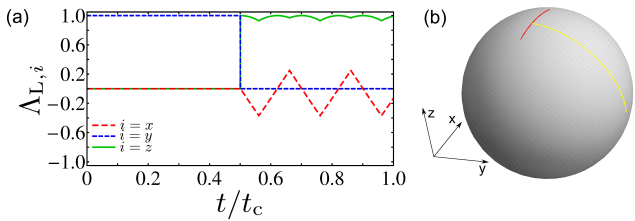}
\caption{(Color online) Components of the time-dependent
spin-polarization in the left well $\Lambda_{\rm{L},i}(t)$
 when the { same driving fields  are used as in}
Fig.~\ref{fig2}. (b) The trajectories on the unit sphere
as determined by the dynamics of  $\vec{\Lambda}_{\rm{L}}(t)$.
 Yellow (red) color corresponds to $t\le 0.5
t_{\rm c}$ ($t> 0.5 t_{\rm c})$.} \label{fig3}
\end{figure}

The results of the corresponding numerical simulation are
illustrated in Fig.~\ref{fig2} where the spin-resolved
time-dependent probability of finding an electron in the left well
[$P_{L\uparrow}(t),P_{L\downarrow}(t)$] is shown. We start from the
tunneling initial condition (both spin states have the same
probability). {After a  time} $t_{0}=0.25t_{\rm c}$ of free
propagation a $x$-HCP with $\beta=\pi/4$ is applied, which
transfers the system into the ground state (both spin states have
the same probability $P_{L\uparrow}=P_{L\downarrow}=0.25$). At
$t=0.5t_{\rm c}$ we apply a $y$-HCP with $\alpha=\pi/4$ and obtain
immediately a nearly perfect spin separation. This means that the
two spin states are localized in different wells:
$P_{L\uparrow}=0.5$ and $P_{L\downarrow}=0$. This separation is
then maintained by applying a periodic $x$-HCP train with the
effective strength $\beta=\pi/2$, period $T=0.1t_{\rm c}$ and the
first HCP centered at $t=0.56t_{\rm c}$. As a result,
{the mean values} of the probabilities after the separation,
averaged over the time period $2t_{\rm c}$ {are}: $\langle
P_{L\uparrow}\rangle=0.495$ and $\langle
P_{L\uparrow}\rangle=0.046$. Thus a very good spin-separation is
stabilized in time. This is also illustrated in Fig.~\ref{fig3}a
where the dynamics of the components of the spin-polarization
$\vec{\Lambda}_{\rm{L}}(t)$ in the left well are shown. Until the
time moment of applying the $y$-HCP at $t=0.5 t_{\rm c}$ the spin
polarization in the left well is oriented along the $y$-axis. Just
after this time moment the spin polarization turns abruptly into
the $z$-direction. Its orientation oscillates then in a small
solid angle close to the $z$-axis under the action of the
maintaining $x$-HCP train. The time-averaged (over $2t_{\rm c}$)
spin polarization in $z$-direction is $0.982$. The corresponding
polarization vector trajectory on the unit sphere is shown in
Fig.~\ref{fig3}b.

Let us discuss shortly the conditions required for an experimental
realization.  As a first choice for the model DQD we may consider
a typical GaAs-based symmetric DQW as in
Ref.~\onlinecite{Alex_QW2004}, {where $t_{\rm c}\approx 0.67$~ps.
We may take $\tau_{\rm d}=40~\rm fs$ ($\tau_{\rm d}$ should be varied in a way that the light
can still drives effectively the relevant dynamics).
 Values of $\alpha_{_{\rm SO}}$ for different}
A$_3$B$_5$ semiconductor materials are listed in
Ref.~\onlinecite{Andrada1997}. For GaAs we have $\alpha_{_{\rm
SO}}\approx 4 ~\mbox{\AA}^2$. Then we can estimate the maximum
required amplitude of HCPs to $E_0\sim 10^7~{\rm V/cm}$
\footnote{These values are  for HCPs polarized in the
$y$-direction, for which the motion is limited by the quantum
confinement. Required amplitudes for $x$-HCPs are two orders of
magnitude lower.}. Such high and short fields have become
available  recently
\cite{Sell2008,Junginger2010} but one may wish to attenuate  them in
order to avoid undesirable effects. {In particular, strong
fields may alter the
 band structure (e.g., via the dynamic Franz-Keldysh or  the AC Stark effects)
 that is not accounted for here.} Narrow gap semiconductors, in particular InSb,
are suitable as a replacement for GaAs because of the much higher
value of $\alpha_{_{\rm SO}}$ \cite{Andrada1997}.

%One of possibilities to maintain the spin separation achieved due
%the application of the y-HCP would be to apply immediately an
%appropriate pulse x-HCPs, that means the single x-HCPs have the
%same pulse strength and occur with fixed time distance $T$. It is
%already shown that an appropriate train of x-HCPs can be used to
%maintain single spin states in a DQD\cite{Alex_QW2004}. The
%strength and duration of the single pulses of the train are in the
%same magnitude as the x-HCP, which was used to generate the
%optical initial condition. As long as the train of x-pulses is
%activated the Spin separation can be maintained.

We have shown that it is possible to separate the spin states of
electrons in a double quantum dot by only two different ultrafast
light pulses. We achieved a nearly perfect spin polarization in the
particular direction which can be maintained for a desired
period of time by applying an additional pulse train. Such
light-induced spin filtering can be realized on a sub-picosecond
time scale that can be of relevance for designing ultrafast
spintronic and spin-qubit devices. Heterostructures based on
narrow gap semiconductors with strong spin-orbit interaction are
good candidates for an experimental demonstration of the predicted
phenomena.

J.B. thanks B.Y. Sun for enlightening discussions.
%\bibliography{reportJ}
\end{document}